# First Real-Time Detection of Ambient Backscatters using Uplink Sounding Reference Signals of a Commercial 4G Smartphone

Ahmed ElSanhoury , Islam Galal, Khaled AlKady, Aml ElKhodary, Dinh-Thuy Phan-Huy , *Senior Member, IEEE*

Ayman M. Hassan

*Abstract*— **Recently, cellular Ambient Backscattering has been proposed for cellular networks. Up to now an Ambient backscatter device, called zero-energy device or tag, broadcasted its message by backscattering ambient downlink waves from the closest Base Station (BS) according to a predefined pattern. A tag was detected by smartphones nearby. This paper presents, for the first time, a novel ambient backscatter communication system exploiting uplink ambient waves from smartphones instead of downlink waves. In this novel system, a BS connected to a smartphone monitors the uplink pilot signals and detects TAGs in proximity. The proposed system is implemented and tested with one prototype of TAG, a commercial off the shelf 4G smartphone and a 4G Software Defined Radio (SDR) BS. Indoor and outdoor experiments were conducted to assess the proposed technique. These very preliminary experiments exhibit a promising potential. In indoor, a detection probability of more than 90% has been achieved without false alarm when the TAG was 3 meters from the UE, and the BS 20 meters away of them, behind walls and obstacles.**

*Index Terms*— **LTE, Uplink, Sounding Reference Signals, Backscatter Communication, ZED Detection, Wireless Communication, IoT, Zero Energy Device, ZED.**

## I. INTRODUCTION

Backscattering devices are currently being introduced in cellular networks, under the naming of "Ambient Internet-of-Thing" (A-IoT) devices, in the upcoming releases of the 5th Generation (5G) network standard [1-3]. At a longer time horizon, zero-energy devices (ZED), also using backscattering, are being studied in research on the future 6th generation (6G) networks [4-5]. A sub-category of ZED called crowd-detectable ZED (CD-ZED) has been introduced in [6]. In [6], a CD-ZED backscatters ambient waves from a base station (BS) and is read by a user equipment (UE), i.e. a smartphone, nearby, the later being currently connected to and listening to the BS, in the downlink (DL). In [6] the same CD-ZED also backscatters ambient waves from a UE nearby, and is read by the BS, the later being currently connected to and listening to the UE, in the uplink (UL). The CD-ZED applies to the particular case of mobile networks the principle of ambient backscattering discovered in [7]. Indeed, in ambient backscattering communications, devices backscatter ambient waves to communicate. Ambient sources of signals can be either TV towers [7], Wi-Fi, LoRa, Bluetooth low energy [8][9], BSs from cellular networks [6] or UEs [6]. In [6] the CD-ZED harvests light energy to power its electronic circuits (mainly composed of a microcontroller, an RF switch and an energy harvesting and energy management system) and reach full-time energy-autonomy [6]. CD-ZED are low cost and low power and are detectable by UEs connected to BS, or by BSs connected to UEs. They can advantageously be used to develop IoT services almost "out of thin air", i.e. without deploying additional infrastructures.

This work is in part supported by the European Project Hexa-X II under (grant 101095759). *(Corresponding author: Ahmed ElSanhoury)*.

Ahmed ElSanhoury is with Orange Innovation Centre, Orange Egypt, Giza, 12578 Egypt (e-mail: ahmed.elsanhoury@orange.com).

Islam Galal is with Orange Innovation Centre, Orange Egypt, Giza, 12578 Egypt (e-mail: islam.galal@orange.com).

Khaled Alkady is with Orange Innovation Centre, Orange Egypt, Giza, 12578 Egypt (e-mail: khaled.alkady@orange.com).

Aml ElKhodary is with Orange Innovation Centre, Orange Egypt, Giza, 12578 Egypt (e-mail: aml.elkhodary.ext@orange.com).

Dinh-Thuy Phan-Huy is with Networks Department, Orange Innovation, Châtillon, Paris, 92356 France (e-mail: dinhthuy.phanhuy@orange.com).

Ayman M. Hassan is with the Department of Electrical Engineering, Benha University, Benha, Al-Qalyubia 6470015 Egypt, and also with Orange Innovation Centre, Orange Egypt, Giza, 12578 Egypt (e-mail: ayman.mohamed@bhit.bu.edu.eg, ayman.hassan@orange.com).

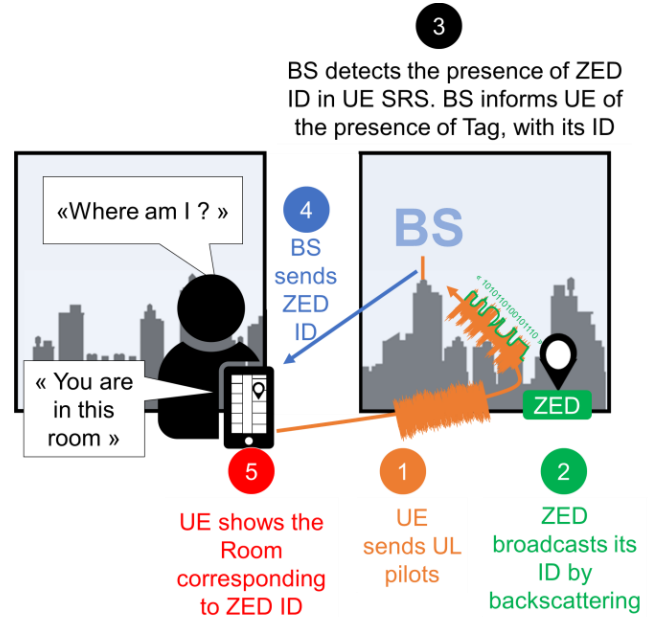

**Fig. 1.** Example of use case of indoor positioning of UE thanks to CD-ZED and UL ambient backscattering illustrated

through several steps: (1) UE sends UL pilots; (2) ZED backscatters the UE signal to broadcast its identity (ID); (3) the BS detects the ZED message in the received UL signal from the UE; (4) the BS informs the UE with the ZED ID, (5) the UE deduces it is currently in the room corresponding to the ZED ID.

In terms of concrete application, CD-ZED can be put on objects to track their locations approximatively and without any additional infrastructure deployment [6]. In this case, each time a CD-ZED gets close to a UE that is connected to the mobile network, geo-localized through Global Positioning System (GPS) or Galileo, the message of the CD-ZED is detected either by the UE (thanks to DL ambient backscattering) or by the BS in signal coming from the UE (thanks to UL ambient backscattering). The position of the CD-ZED is then approximated to be the one of the UE.

Another concrete application that is foreseen, is indoor positioning of UEs without the deployment of additional infrastructure. In this case, a CD-ZED is put on the wall, in each room of a building and used as wireless beacon [10]. As the CD-ZED location is fixed, known and reported on a map, each time a UE gets close to the CD-ZED of a room, it detects the CD-ZED (thanks to DL ambient backscattering) or the BS detects the CD-ZED in the signal coming from the UE (thanks to UL backscattering).

Fig. 1, illustrates this indoor positioning use case, in the case where UL backscattering is used.

Regarding performance, CD-ZED prototypes have first been tested with 4G and 5G ambient downlink signals [10]. The obtained detection performance was good when the mobile network was fully loaded with strong and stable downlink ambient signal. However, the performance collapsed when the mobile network data traffic was bursty (which is the case most of the time in a real commercial mobile network). Indeed, the prototyped UE was not able to distinguish between time-variations (of the received DL ambient signal) due to the CD-ZED from the ones due to the ambient network data traffic. To overcome this limitation, [11-12] have designed a novel detector that listens to DL pilots only instead of listening to the full pilot plus data downlink signal, as pilots are stable over time. This was tested successfully with ambient 4G commercial networks with real bursty data traffic [13-14]. A further detection performance improvement has been obtained by adding channel coding [15].

Up to now, to our best knowledge, UL ambient backscattering in mobile networks has never been studied, implemented nor tested, and its performance on the field remain unknown. In this paper, for the first time, we present a complete UL ambient backscattering system for mobile network test-bed, including prototypes of CD-ZED, commercial 4G UEs and a software defined radio (SDR) base station, the latter one detecting the CD-ZED in real time.

The following sections are organized as follows: Section II describes the end-to-end system architecture; Section III, the ZED detection technique implemented at BS is explained. Multiple tags could be identified using the algorithm demonstrated in the IV section. Some signal processing methods are shown in the V section to enhance the tags identification process and overcome wireless channel variations. Finally, experimentation results are revealed, followed by the conclusion in the VI and the VII sections, respectively.

From now all, we will call CD-ZED simply TAG, to simplify notations, and for the comfort of the reader.

## II. COMMUNICATION SYSTEM OVERVIEW

We consider one single UE being connected to a single BS, through the 4G standardized air interface. The frequency division duplex (FDD) bands with [777-787] MHz for the UL and [746-756] MHz for the DL are being used. All the control and data signals exchanged between the UE and the BS are standardized 4G signals. They must therefore be seen as ambient 4G signals.

As illustrated by Fig. 2, we introduce a single TAG. The TAG communicates its ID to the BS, by backscattering UL ambient 4G signals generated by a single UE, according to the following steps:

- Step#0: the BS configures the UE for periodical sending of UL standardized pilots called sounding reference signals (SRS), using a standardized radio resource control (RRC) DL signaling message [16];
- Step#1: periodically, the UE sends UL SRS, following the 4G standard;
- Step#2: the TAG sends a message (its own ID) continuously and repeatedly by backscattering the UE UL SRS;
- Step#3: the BS listens to the UE SRS, following the 4G standard. In addition, it searches for the TAG message (i.e. the TAG ID) in the received SRS signal.

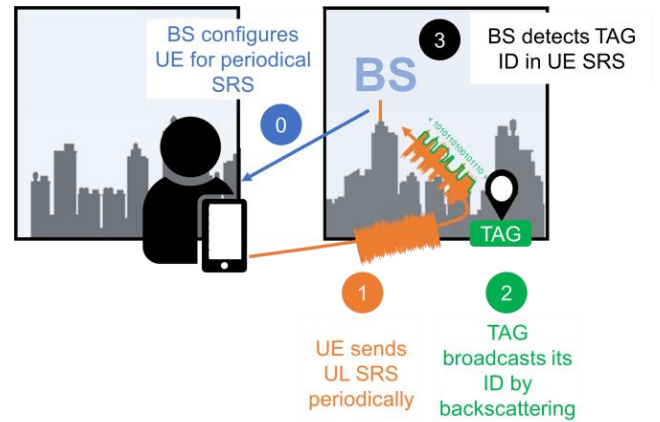

**Fig. 2.** Communication system overview

Each step is detailed hereafter.

### A. Step#0

This step is fully compliant with 4G standard. The SRS is a standardized UL pilot signal sent by the UE, to enable the BS to estimate the UL channel quality between the UE and the BS. The SRS can be sent periodically, and the period is configurable. According to the standard [16], the shortest period is 2 ms, and the largest period is 320 ms. As 4G standard relies on a multi-carrier waveform with frequency division multiplexing, the bandwidth on which the SRS spans is also configurable. Finally, according to the standard, the BS can configure this period and the bandwidth, through an RRC signaling message called "System Information Block Type 2" (SIB2) in DL to the UE. In our system, the periodicity is set to 10ms. The bandwidth is set to 4.32 MHz. This corresponds to parameter *srs-BandwidthConfig* of SIB2 message being set to 5. as illustrated at Fig. 3.

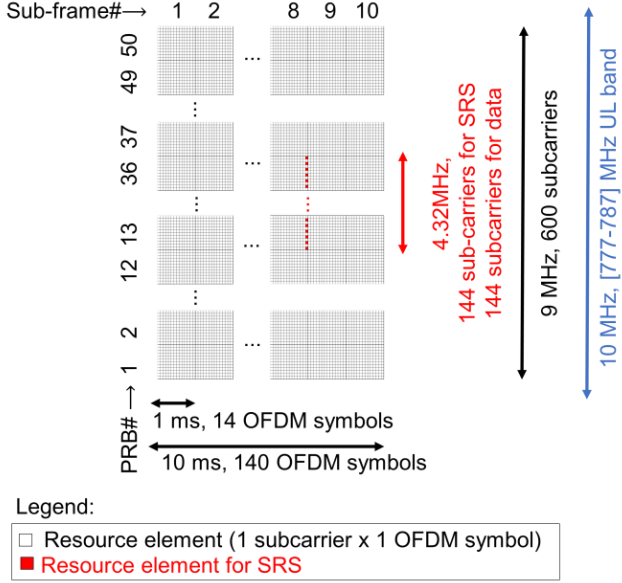

**Legend:**

☐ Resource element (1 subcarrier x 1 OFDM symbol)
■ Resource element for SRS

**Fig. 3.** SRS mapping on time and frequency resources

### B. Step#1

This step is fully compliant with 4G standard. The UE transmits the SRS (according to the received RRC SIB2 message sent during step#0) periodically, every 10 ms over a band of 4.32 MHz, and following the 4G standard [17]. More precisely, as illustrated by Fig. 3, in the time domain, the SRS is mapped on the last orthogonal frequency division multiplex (OFDM) symbol of the $8^{th}$ sub-frame (SF) of 1 ms. In the frequency domain, the SRS is mapped over 144 sub-carriers around the central carrier frequency at 782 MHz. There is one data or empty sub-carrier between each SRS sub-carrier. Equivalently, the SRS is mapped over 24 consecutive physical resource blocks (PRB) of 180 kHz (PRB#13 to 36) among the 50 PRBs. Finally, as specified by the 4G standard [17], the SRS is a Zadoff-Chu Sequence (ZCS) of 144 complex numbers. Note that the ZCS have a constant magnitude and are sent with a constant power. Hence, they share the same good properties as the downlink pilots in LTE, i.e. being stable over time. Hence, as DL pilots in [11-13], these UL pilots are good ambient signal candidates for TAG detection.

### C. Step#2

This step does not exist in 4G standard and is new. As in [6], the TAG is constituted of a dipole antenna that switches between two different states, thanks to an RF switch. In one state, the two branches of the dipole are short-circuited, to make the antenna resonate at the frequency of the ambient 4G, and backscattering 4G waves. In the other state, the antenna is transparent to ambient waves. One state corresponds to bit "0" and the other state corresponds to bit "1". Hence, the TAG sends a binary sequence by switching between states, using On Off Keying (OOK). $T_s$ is chosen equal to the SRS periodicity. Hence $T_s$ = 10ms. This is necessary to ensure the success of step#3.

The binary sequence sent by the TAG is unique, as it also represents the ID of the TAG. We build the sequence as the repetition of one Gold Code. Gold Codes are pseudo-noise sequences generated by XORing two maximal-length sequences, ensuring very low cross-correlation characteristics. Therefore, they are commonly used in code-division multiple access applications that require quasi-orthogonal codes. For our particular system, we use a Gold Code to minimize false-positive detection and to ensure a reliable differentiation between several different TAGs. The two best polynomials to generate a 31-length Gold Codes have been chosen:

$$x^5 + x^2 + 1 \tag{1}$$
$$x^5 + x^4 + x^3 + x^2 + 1 \tag{2}$$

The TAG repeats each bit of the Gold sequence several times. Let $N \in \mathbb{N}$ be the length of the Gold sequence transmitted by the TAG and $v \in \mathbb{N}$ the number of repetitions, we denote by $\mathbf{x} \in \{-1,1\}^{N \times 1}$ the original Gold sequence and we denote by $\mathbf{x}' \in \{-1,1\}^{vN \times 1}$ the transmitted message. Hence, for each $1 \le n \le N$, and each $1 \le q \le v$, we have:

$$\mathbf{x}'_{q+(n-1)v} = \mathbf{x}_n \tag{3}$$

We chose N = 31 and v = 7, as a result, the duration of the TAG message Tm is $T_m = N \times v \times T_s = 31 \times 7 \times 0.001 = 2.17$ s, which is high, and limits the usage of the studied system to static UEs, i.e. typically smartphones of users sat in a meeting room.

Fig. 4 illustrates the Gold code sent by the TAG and the full set of codes (each of the two polynomials + 31 together) among which this code must be searched by the BS.

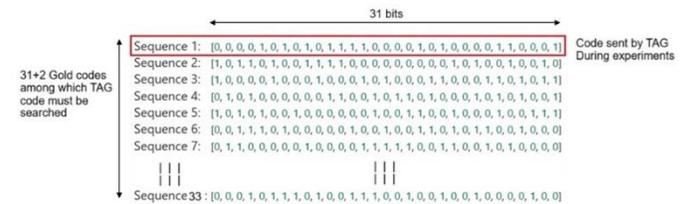

**Fig. 4.** Gold Codes used in the system

### III. DEEP-DIVE ON THE TAG DETECTOR

This section details step#3, i.e. the signal processing at the BS side to detect the ZED. As in [11-13], we propose to avoid using the full UL ambient signal, and only listen to the UL pilots, i.e. the SRS. As illustrated by Fig. 5, step#3 is composed of several consecutive blocks:

1. SRS extraction block;
2. magnitude averaging block;
3. filtering block;
4. correlation of with Gold Code block;
5. TAG ID detection.

Sub-steps #1 and #2 already exist in commercial BSs. Sub-steps #3 and #4 would need to be added, for TAG detection.

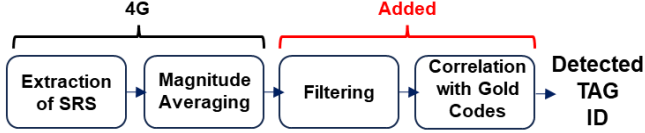

**Fig. 5.** ZED detector at BS side

The following sub-sections detail each block.

### A. Magnitude averaging block

In the magnitude averaging block, the BS extracts the SRS every $T_s$. Let $k \in \mathbb{N}$ be the current $T_s$ period. The result of the SRS extraction is the complex vector $\mathbf{s}^{(k)} \in \mathbb{C}^{N \times 1}$ where $N = 144$. The average magnitude $a^{(k)} \in \mathbb{R}^+$ for $T_s$ period number $k$ is computed by the BS as follows:

$$a^{(k)} = \frac{1}{N} \sum_{n=1}^{N} \left( |\mathbf{s}_n^{(k)}| \right). \tag{4}$$

The output of the averaging block is $\mathbf{x}^{(k)}$.

### B. Filtering block

The filtering block aims at removing amplitude spikes due to our SDR set-up and hardware impairments. The filtering block is also useful to account for sudden changes due to UE move or change of orientation.

Our filtering has three successive sub-blocks. In the first sub-block, we set a hard threshold $\alpha \in \mathbb{R}^+$ to check the validity of the SRS magnitude. If an invalid value is detected, it is replaced using the last valid SRS signal after all filtering processes. We denote $a'^{(k)} \in \mathbb{R}^+$ the output of this sub-block at $T_s$ period $k$. $a'^{(k)}$ is obtained as follows: if $a^{(k)} > \alpha$, then $a'^{(k)} = a'^{(k-1)}$, $\mathbf{x}'^{(k)} = \mathbf{x}^{(k)}$ otherwise (with $a'^1 = a^1$). Note that in a commercial BS (without hardware impairments due to our SDR set-up), this stage would not be necessary. Based on our hardware and SDR set-up, we set the value of $\alpha$ to 0.55.

In the second sub-block, a "median filter" is used to remove sudden strong fluctuations (spikes due to hardware impairments and SDR). A sliding window approach with depth P is used. To ensure that the valid gold code sequence and ZED-induced modulation remain intact, the depth of the median filter P is kept shorter than the Gold Code repetition period. This prevents excessive smoothing that could remove legitimate variations. Taking into account these constrains, we set the value of P to 5 samples (i.e. 50ms). More precisely, a buffer of length P is fed with the last P outputs of the previous sub-block. Let $\mathbf{b}^{(k)} \in \mathbb{R}^{+^{P \times 1}}$ be the buffer during the k-th $T_s$ period. $\mathbf{b}^{(k)}$ is given by: $\mathbf{b}^{(k)} = [\mathbf{x}'^{(k-P+1)} \ ... \ \mathbf{x}'^{(k)}]$. Let $\mathbf{c}^{(k)} \in \mathbb{R}^{+^{P \times 1}}$ be a sorted copy $\mathbf{b}^{(k)}$, where the elements of $\mathbf{c}^{(k)}$ are the same as the elements of $\mathbf{b}^{(k)}$, with the difference that there are sorted in increasing order. Let $d'^{(k)} \in \mathbb{R}^+$ be the output of the median filter at Ts period $k$. $d^{(k)}$ is computed as follows:

$$d^{(k)} = \mathbf{c}_{k-P+1+P/2}^{(k)} \text{ if P is odd and} \tag{5}$$
$$d^{(k)} = \mathbf{c}_{k-P+1+\frac{P+1}{2}}^{(k)} \text{ if P is even.}$$

Note that again, in a commercial BS (without hardware impairments due to our SDR set-up), this stage would not be necessary. More precisely, a buffer of length P is fed with the last P outputs of the averaging block.

In the third sub-block, a "standard deviation" (SD) filter with a window size Q smaller than the Gold Code repetition period is used to detect and correct remaining outliers, enhancing overall signal stability. Taking into account these constrains, we set the value of Q to 5. More precisely, a buffer of length Q is fed with the last Q outputs of the previous sub-block. Let $\mathbf{e}^{(k)} \in \mathbb{R}^{+^{Q \times 1}}$ be the buffer during the k-th $T_s$ period. $\mathbf{e}^{(k)}$ is given by: $\mathbf{e}^{(k)} = [d^{(k-Q+1)} \ ... \ d^{(k)}]$. The mean $\varepsilon^{(k)}$ and the standard deviation $\sigma^{(k)}$ of the samples in $\mathbf{e}^{(k)}$ are computed. Let $y^{(k)}$ be the output of this sub-block. If $\left| \mathbf{e}_Q^{(k)} - \varepsilon^{(k)} \right| > u \times \sigma^{(k)}$, then $y^{(k)} = m^{(k)}$, $y^{(k)} = \mathbf{e}_Q^{(k)}$ otherwise, where $u$ is a deviation factor set to 20% to track channel variations.

### C. Correlation with Gold codes block and TAG ID detection

In this block, the output of the filtering block is correlated with all possible Gold Codes. If the correlation value for one given Gold Code exceeds a predefined threshold $\theta$, the TAG is considered as detected, and its ID is the detected Gold Code. The three commonly used correlation methods are explored: Pearson, Spearman, and Kendall [18]. Pearson's coefficients measure linear correlation, while Spearman and Kendall focus on ranking-based comparisons. Kendall's correlation, with $O(n^2)$ complexity, is computationally expensive, and Spearman's $O(n \log n)$ complexity introduces additional overhead due to ranking operations. In contrast, Pearson's $O(n)$ complexity ensures faster execution and lower processing overhead, making it ideal for efficient real-time detection [19]. After exploring all three methods, only Pearson's method is retained, as it demonstrated the fastest computation time, making it the most suitable choice for real-time TAG detection. The Pearson's correlation $r^{(k)}$, is computed as follows at $T_s$ period k:

$$r^{(k)} = \frac{\sum_{n=1}^{vN} (\mathbf{x}'_n - \mu) \left( y^{(k+n-vN)} - \rho^{(k)} \right)}{\sqrt{\sum_{n=1}^{vN} (\mathbf{x}'_n - \mu)^2 (y^{(k+n-vN)} - \rho^{(k)})^2}} \tag{6}$$

where $\mu = \frac{1}{vN} \sum_{n=1}^{vN} \mathbf{x}'_n$ and $\rho^{(k)} = \frac{1}{vN} \sum_{n=1}^{vN} y^{(k+n-N)}$.

If $\rho^{(k)} > \theta$, then, the TAG is considered as detected, and its ID is given by the Gold Code $\mathbf{x}$ on which $\mathbf{x}'$ is based. As the BS searches for one sequence among several candidates, the test is performed for all possible sequences introduced in Section II, and the sequence with highest correlation is picked, as illustrated by Fig. 6. This exploits the good autocorrelation and cross-correlation properties of the Gold codes, as illustrated for two particular Gold codes in Fig. 7. In our system, θ is chosen equal to 0.4.

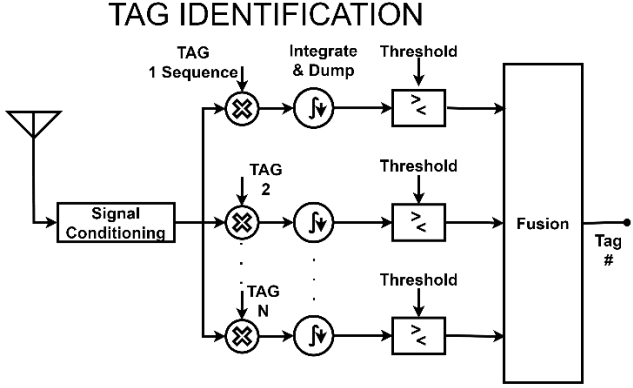

**Fig. 6.** TAG ID detection.

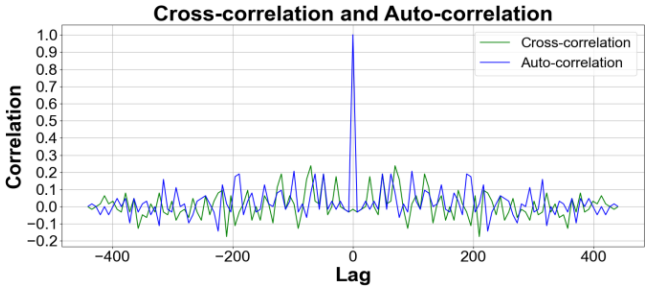

**Fig. 7.** Autocorrelation versus Cross-correlation

## IV. Experimental Set-Up and Evaluation Methodology

### A. Set-Up

The system described in Sections II and III is implemented in a real time test-bed depicted in Fig. 8. where, the TAG is a prototype, the UE is a Commercial Off-The-Shelf (COTS) Smartphone, the Samsung S8, namely, and the BS is a Software Defined Network (SDN) platform connected to the Internet.

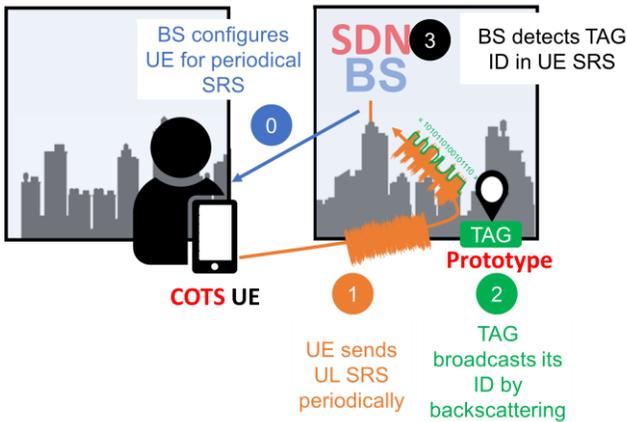

**Fig. 8.** Test-bed composed of a COTS UE, a prototype TAG and an SDN BS.

Fig. 9 illustrates the SDN platform, which is composed of an USRP B210 connected to a laptop on which srsRAN [20] is run, as a standard 4G BS and a standard core network is run. The laptop is connected to the internet. Therefore, the COTS UE is connected to the internet through the SDR BS. In the srsRAN software, we have reused the srsEPC block (corresponding to the 4G core network) without any modification and we have modified the eNode B software (corresponding to the 4G BS) to implement step#2 and step#3 of our system, as presented in Sections II and III.

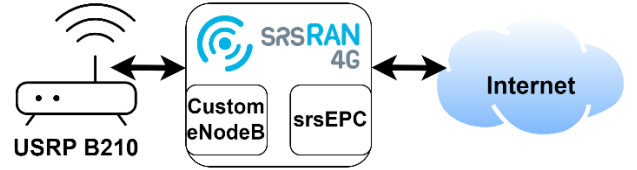

**Fig. 9.** SDN platform composed of srsRAN running on a laptop and a USRP B210, connected to the Internet.

Regarding the RF part of the BS, we have used the antenna DeLOCK 88571 (with 3 dBi of antenna gain). The antenna is connected through the frequency duplexer DCM751-782-10A1 operating on Band 13 frequencies [21], to two branches: the transmitter branch and the receiver branch of the BS. On the transmitter branch we use a power amplifier (PA) and an attenuator (Att), to control the transmit power of the BS. On the receiver branch, we use a Low Noise Amplifier (LNA). Additionally, a bandpass filter operating in the range of 768-795 MHz is used.

Several different RF configurations have been used depending on the test scenario. Configuration A, illustrated in Fig. 10, was used in indoor for short BS-UE ranges, whereas Configuration B, illustrated in Fig. 11 and Fig. 12, was used for outdoor tests and larger UE-BS ranges (with strong PA and LNA).

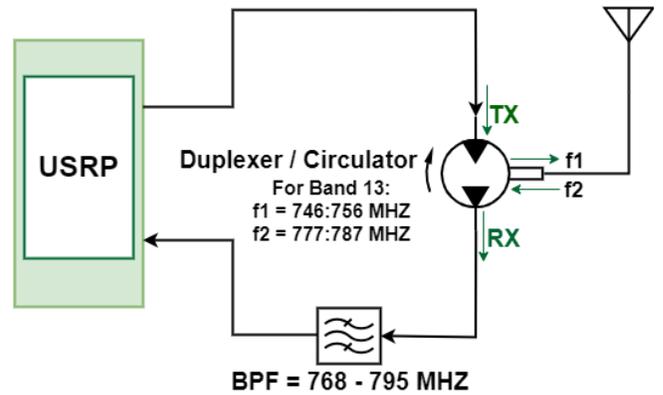

**Fig. 10.** RF configuration A, for indoor tests (block diagram view).

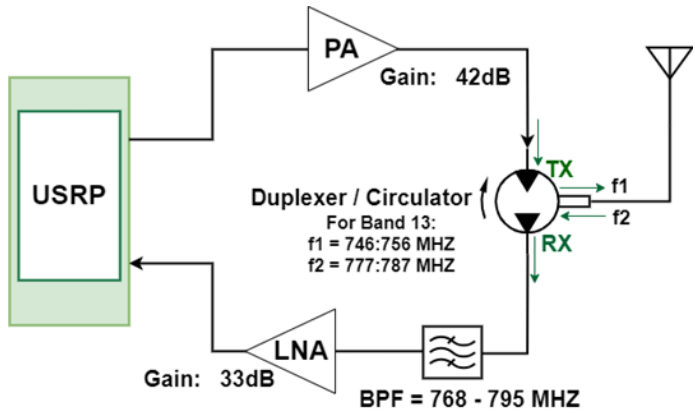

**Fig. 11.** RF Configuration B for outdoor tests (block diagram view).

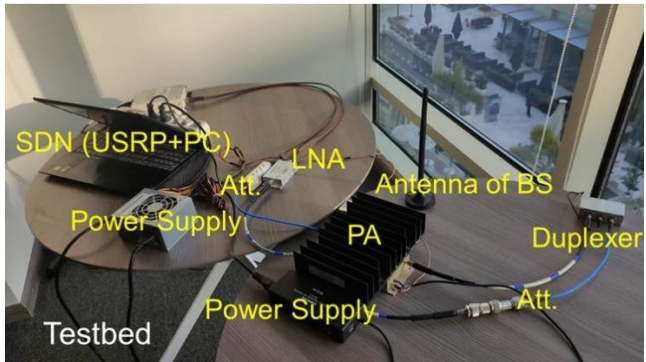

**Fig. 12.** RF Configuration B for outdoor tests (implementation view)

### B. Evaluation methodology

The flowchart of the protocol we implemented, as illustrated in Fig. 13, has four phases:

- The "attached phase", which is 4G standard compliant, and during which, step#0 from Section II is performed;
- The "TAG OFF" phase, which is not necessary in the System described in Section II and III. We introduce to perform some SRS measurements when the TAG is not switching and applying OOK;
- The "TAG ON" phase, which is not necessary in the System described in Section II and III. We introduce to perform some SRS measurements when the TAG is switching and applying OOK;
- The "BS Processing phase", which implements the TAG detection depicted in Section III.

Fig. 14 details the BS processing phase. We detail hereafter how the communication system performance is measured during the BS processing phase. For each $T_s$ period, the detection event is collected. For a given UE, TAG and BS position, the detection probability, the false alarm probability and the cross false alarm probability are measured. The cross false alarm is defined as follows: this event occurs when the TAG is powered ON but is detected with a gold code that is not pre-installed on the specified TAG. The probabilities are computed thanks to the sending of R=300 times the same sequence by the TAG. One measurement therefore lasts for the duration $T_{meas} = R \times Tm = R \times v \times N \times T_s = 300 \times 2.17 = 650\ s = 10\ min\ 51\ s$. One measurement relies on the sending of $Nsrs = R \times v \times N = 65100$ SRS sequences. These long measurement periods are necessary to collect enough data to get statistical results.

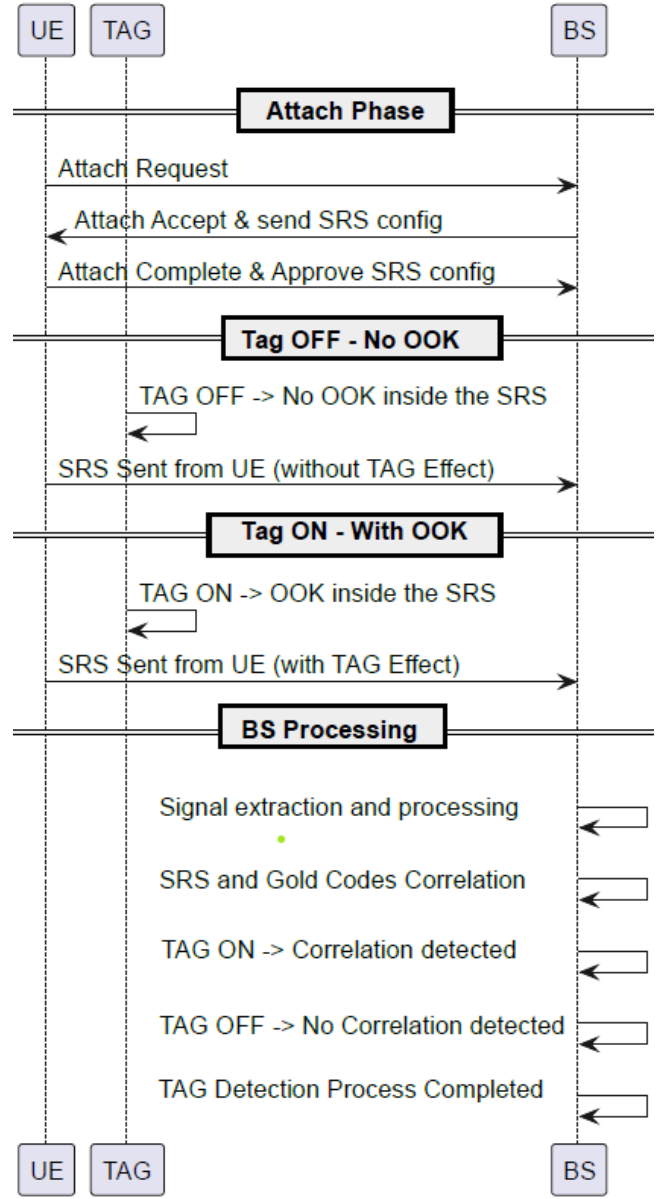

**Fig. 13.** Signal Flow of TAG Detection

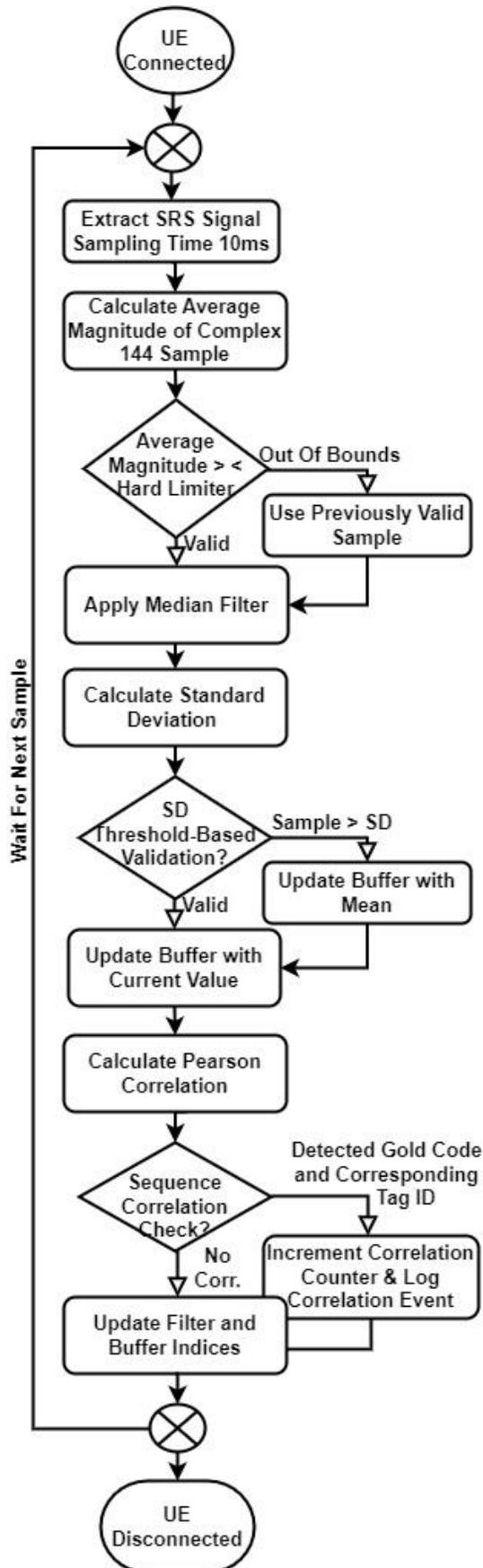

**Fig. 14.** BS processing phase detailed, including detection event collection.

## V. RESULTS

Experiments have been run with the set-up and methodology presented in Section IV. Note that as only R=300 samples have been used for each probability measurement, these numbers must be considered as rough approximates of the true probabilities.

### A. Measurements during TAG ON and TAG OFF Phase

This section analyses results obtained during the "TAG ON" and "TAG OFF" phases. We recall that these phases are not part of the communication system described in Sections II and III, and would therefore not be needed in a real system. These phases have been added for the purpose of measurement. During each of these two phases, SRS signals have been sent by the UE and measured at the BS side. The average magnitude (the $a^{(k)}$ symbols of Section II) of the SRS signals is plotted as a function of time, for each phase in Fig. 15, for a duration of around 10 seconds. One can observe that when the TAG is "OFF", the $a^{(k)}$'s are almost constant, with minor deviations only due to surrounding external effects. On the contrary, when the TAG is "ON", and switches to apply OOK, the $a^{(k)}$'s evolve over time. We also observe strong short spikes, which are not usually seen in previous experiments we have conducted or participated to [6], [11-15]. We believe these are due to our SDR setup and hardware impairments. The first sub-blocks of the filter block described in Section III-B), have been designed to clean the signal from these unwanted spikes. In a commercial BS such spikes may not appear.

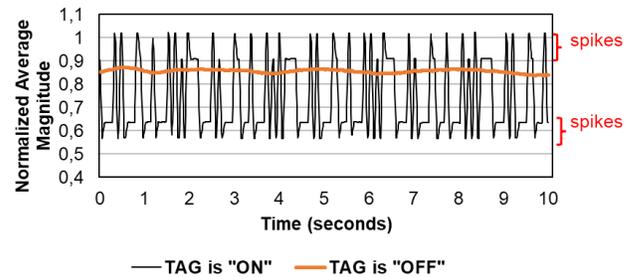

**Fig. 15.** Output of the SRS magnitude averaging

### B. Measurements in Scenario#1: Indoor short range

In this first scenario, we considered a controlled environment inside a room. The distance between the BS and the TAG was 6 meters, and the TAG-to-UE distance was fixed to 1 meter. RF Configuration A described in Section IV-A) has been used. Fig. 16 illustrates Scenario#1 deployment.

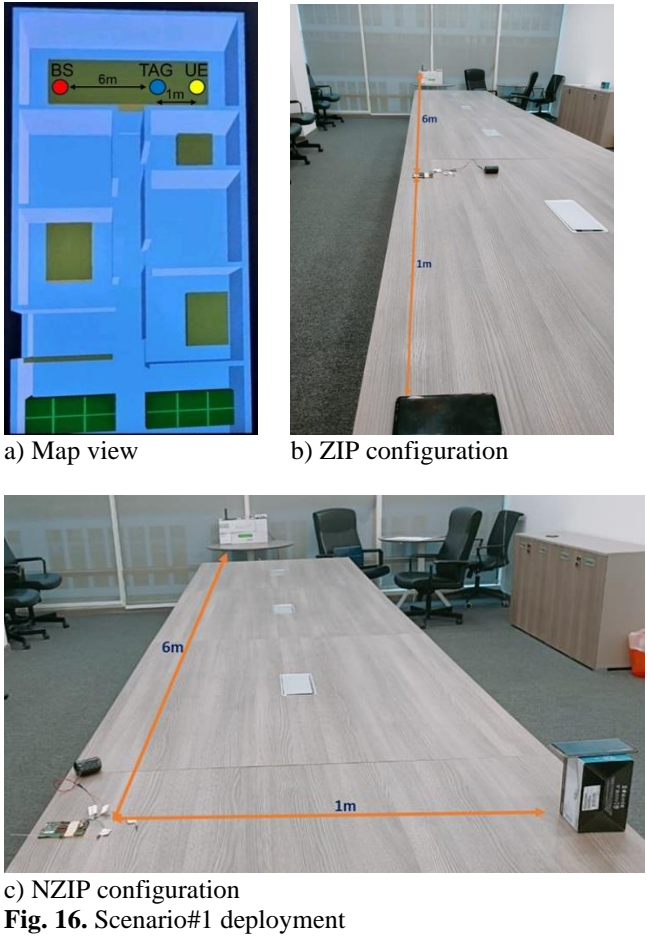

a) Map view     b) ZIP configuration

c) NZIP configuration
**Fig. 16.** Scenario#1 deployment

We measured the system's ability to maintain a strong correlation between the SRS and the gold code under varying orientations of the UE at close distances. The objective was to measure to which extent correlation remained consistent despite changes in UE inclination with respect to the TAG. More precisely, we define the following two orientations:

- Zero-Inclination Position (ZIP): UE is directly aligned with the TAG dipole, meaning the TAG faces the mobile.
- Non-Zero Inclination Position (NZIP): UE is rotated ±90 degrees relative to the TAG.

The results in Table I indicate that the TAG detection exceeded 90%, and the false alarm remained null, whatever the UE inclination relative to the TAG.

TABLE I
DETECTION RESULTS VS ORIENTATION (OVER R=300 SAMPLES)

| Orientation | Detection | False Alarm |
|---|---|---|
| **ZIP** | 98.83% | 0.00% |
| **NZIP** | 92.17% | 0.00% |

*C. Measurements in Scenario#2: Indoor long range*

In this second scenario, still in indoor, we tested longer BS-UE distances, reaching up to 20 meters. The testing area included obstacles and walls between the BS and the UE, which introduced significant fading. The detection probability, the false-alarm probability and the cross false alarm, at different UE orientations relative to the TAG (ZIP and NZIP) are evaluated. For this scenario, RF Configuration A defined in Section IV-A) has been used. Fig. 17 illustrated Scenario#2 deployment.

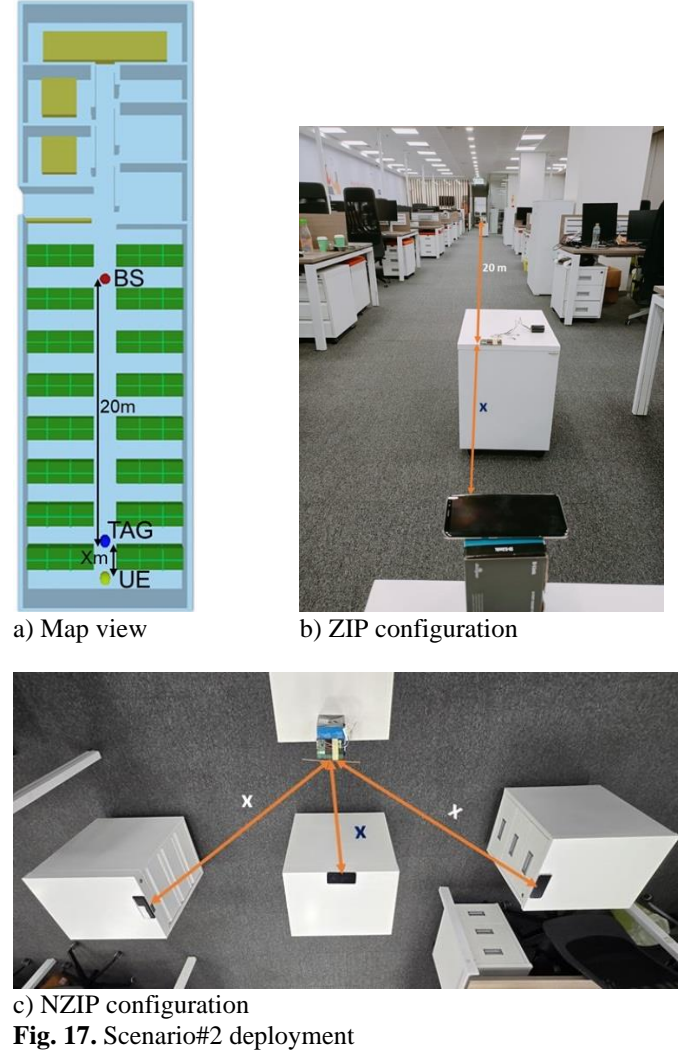

a) Map view     b) ZIP configuration

c) NZIP configuration
**Fig. 17.** Scenario#2 deployment

In 90% of measurements, the false alarm was 0%. While in 95% of measurements, the false alarm was greater than 0%, and the maximum value was 1%.

The detection probability is reported in Fig. 18. We observe that for NZIP configuration, whatever the UE-TAG distance, the detection probability remains high. On the contrary, for ZIP configuration, when the EU-TAG distance exceeds 3 meters, the detection probability collapses.

The cross false alarm is reported in Table II. The results indicate that this type of false alarm does not exceed 1%, as shown in the table, even when the distance reaches up to 5 meters.

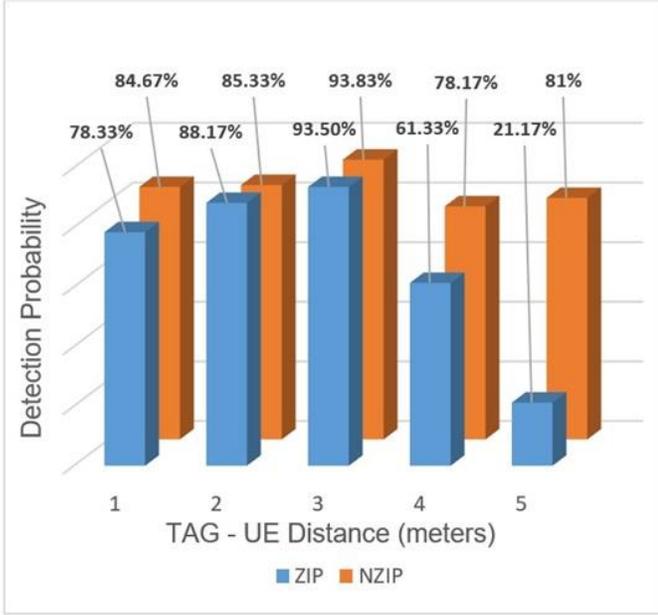

**Fig. 18. TAG** Detection Results for Scenario#2

TABLE II
Cross False Alarm Results Vs Orientation

| UE & TAG distance in m | Cross False Alarm | |
|---|---|---|
| | ZIP | NZIP |
| 1 | 0.5% | 0.00% |
| 2 | 0.00% | 0.00% |
| 3 | 0.00% | 0.00% |
| 4 | 0.33% | 0.5% |
| 5 | 0.67% | 1% |

It is noticeable that with an indoor UE-BS distance of 20 meters, in NLOS, and a TAG-UE distance of 3 meters, the detection probability exceeds 90% and without false alarm.

*B. Measurements in Scenario#3: Outdoor-to-indoor*

In this scenario, we tested more challenging conditions, with indoor-to-outdoor communication from the UE outdoor the BS indoor. For this scenario, RF Configuration B defined in Section IV-A) has been used. The distance was extended to 60 meters in both ZIP and NZIP scenarios.

TAG detection was successful, although the correlation scores slightly decreased due to channel conditions. The measured detection probability was 30% with 5% false alarm using NZIP configuration between UE and TAG at outdoor. Test setup of scenario#3 is illustrated at Fig. 19.

Future enhancements, such as AI-driven channel estimation and adaptive correlation techniques [22], could further improve detection accuracy by dynamically adjusting filtering parameters based on real-time environmental conditions.

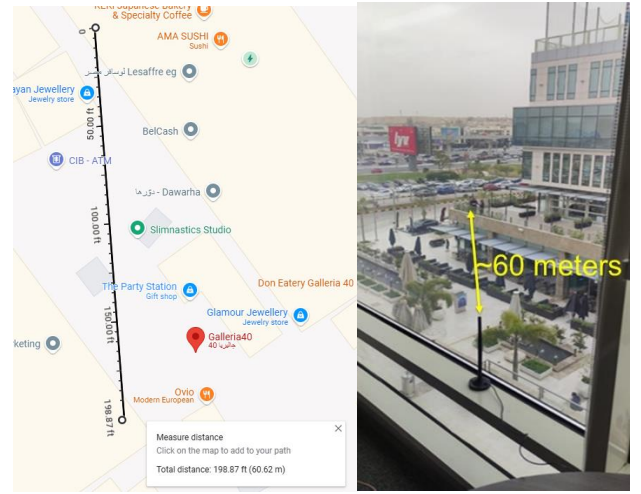

**Fig. 19.** The Outdoor distance between TAG (and UE) and eNodeB

## VII. Conclusion and Future Work

A novel backscattering communication technique is presented and proved experimentally. BD or TAG can communicate with its reader (LTE BS) piggybacked on the LTE UL signal using COTS UE. Through the SRS, the COTS UE is commanded and configured during its attach procedure to send SRS regularly. SRS's signal, which is based on the Zadoff–Chu sequence, is exploited to reflect the effect of the pseudo-noise pattern generated by the BD when the magnitude of the SRS differs from one.

Practical indoor/outdoor experiments were conducted to assess the proposed technique. These very preliminary experiments show that communication seems feasible for different scenarios in indoor and outdoor. In indoor, a detection probability of more than 90% has been achieved without false alarm when the TAG was 3 meters from the UE, and the BS 20 meters away, behind walls and obstacles.

Future work will focus on enhancing detection parameters and mitigating channel variations using AI-driven adaptive filtering and improved channel estimation techniques. Using AI-empowered techniques discussed in the literature, improving signal detection, channel estimation, and interference mitigation could be investigated.

## References

[1] "Study on Ambient IoT (Internet of Things) in RAN" *3rd Generation Partnership Project (3GPP) Technical Report (TR) 38.848 v18.0.0,* 29th Sept. 2023. Available at: https://www.3gpp.org/ftp/Specs/archive/38_series/38.848/38848-i00.zip.

[2] "Study on Ambient power-enabled Internet of Things" *3GPP TR 22.840 v19.0.0,* 12th dec. 2022. Available at: https://www.3gpp.org/ftp/Specs/archive/22_series/22.840/22840-j00.zip.

[3] "Study on solutions for Ambient IoT (Internet of Things) in NR", *3GPP TR 38.769 v2.0.0,* 4th dec. 2024. Available at: https://ftp.3gpp.org/Specs/archive/38_series/38.769/38769-200.zip.

[4] D. Sabella *et al.,* "Innovation Management in 6G Research: The Case of Hexa-X Project," *IEEE Communications Magazine,* vol. 62, no. 2, pp. 142-149, February 2024.

[5] S. Naser, L. Bariah, S. Muhaidat and E. Basar, "Zero-Energy Devices Empowered 6G Networks: Opportunities, Key Technologies, and Challenges," IEEE Internet of Things Magazine, vol. 6, no. 3, pp. 44-50, September 2023.

[6] D. -T. Phan-Huy, D. Barthel, P. Ratajczak, R. Fara, M. d. Renzo and J. de Rosny, "Ambient Backscatter Communications in Mobile Networks: Crowd-Detectable Zero-Energy-Devices," *IEEE Journal of Radio Frequency Identification*, vol. 6, pp. 660-670, 2022.

[7] V. Liu, A. Parks, V. Talla, S. Gollakota, D. Wetherall, and J. R. Smith "Ambient backscatter: Wireless communication out of thin air," *ACM SIGCOMM computer communication review*, 43(4), pp. 39-50, 2023.

[8] N. Van Huynh, et al. "Ambient backscatter communications: A contemporary survey," *IEEE Communications surveys & tutorials*, 20.4, pp. 2889-2922, 2018.

[9] W. Wu et al. "A survey on ambient backscatter communications: Principles, systems, applications, and challenges." *Computer Networks*, 2022.

[10] S. Yang, Y. Bénédic, D.-T. Phan-Huy, J.-M. Gorce, and G. Villemaud "Indoor Localization of Smartphones Thanks to Zero-Energy-Devices Beacons," in *Proc. 2024 18th European Conference on Antennas and Propagation (EuCAP)*, pp. 1-5, 2024.

[11] K. Ruttik, X. Wang, J. Liao, R. Jäntti and P. -H. Dinh-Thuy, "Ambient backscatter communications using LTE cell specific reference signals," in *Proc. 2022 IEEE 12th International Conference on RFID Technology and Applications (RFID-TA)*, Cagliari, Italy, 2022, pp. 67-70.

[12] J. Liao, X. Wang, K. Ruttik, R. Jäntti and D. -T. Phan-Huy, "In-Band Ambient FSK Backscatter Communications Leveraging LTE Cell-Specific Reference Signals," *IEEE Journal of Radio Frequency Identification*, vol. 7, pp. 267-277, 2023.

[13] P. Ndiaye et al., "Zero-Energy-Device for 6G: First Real-Time Backscatter Communication Thanks to the Detection of Pilots from an Ambient Commercial Cellular Network," *in Proc. 2023 2nd International Conference on 6G Networking (6GNet)*, Paris, France, 2023.

[14] J. Liao et al., "Indoor Backscattering Communication by Using Commercial LTE Pilots," in *Proc. 2024 IEEE 99th Vehicular Technology Conference (VTC2024-Spring)*, Singapore, Singapore, 2024, pp. 1-5.

[15] J. Liao, K. Ruttik, R. Jäntti, M. U. Sheikh and Phan-Huy, D.-T., "Measurement of Coded Backscatter Communication Utilizing Commercial LTE Ambient Signal," in *Proc. 2024 3rd International Conference on 6G Networking (6GNet)*, Paris, France, 2024.

[16] 3GPP TS 36.331 "Radio Resource Control (RRC); Protocol specification," available at: https://www.3gpp.org/ftp/Specs/archive/36_series/36.331/36331-i31.zip, 27th sept. 2024.

[17] "Physical Channels and Modulation (Release 8)" 3GPP TS 36.211 V8.9.0, available at https://www.3gpp.org/ftp/Specs/2024-09/Rel-8/36_series/36211-890.zip, dec. 2009.

[18] M. Shaqiri, T. Iljazi, L. Kamberi and R. Ramani-Halili "Differences Between The Correlation Coefficients Pearson, Kendall And Spearman," *Journal of Natural Sciences and Mathematics of UT 8*, no. 15-16, pp. 392-397, 2023.

[19] W. Xu, C. Chang, Y.S. Hung, S.K. Kwan, P.C.W. Fung "Order statistics correlation coefficient as a novel association measurement with applications to biosignal analysis," *IEEE Transactions on Signal Processing 55 (12)* pp. 5552–5563, 2007.

[20] srsRAN opensource project, https://www.srslte.com/.

[21] "Base Station (BS) radio transmission and reception (Release 18)," 3GPP Technical Specification 36.104 V18.0.0, available at: https://www.3gpp.org/ftp/Specs/archive/36_series/36.104, December 2022.

[22] F. Xu, T. Hussain, M. Ahmed, K. Ali, M. A. Mirza, W. U. Khan and E. Han, Z. "the state of AI-empowered backscatter communications: A comprehensive survey," *IEEE Internet of Things Journal*, 2023.

**Ahmed ElSanhoury**, Ahmed has over two decades of experience in wireless systems and product design. As a lead of IoT Engineering, he has served joint industry-academic activities, strengthening the link between academic research and practical industry needs. On the academic side, after obtaining his B.Sc. and M.Sc. from Cairo University in 1998 and 2006, he is pursuing his PhD in electronics & communications engineering from Cairo University. He published several times at IEEE conferences and presented his work in person overseas in Morocco, Malaysia, and France. Also, he became Informa certified with a Distinction grade from Informa & Derby University in LTE and Advanced Communications. On the industry side, he was designated as an Orange Expert and joined the Orange Experts community in 2020.

**Islam Galal**, Islam received his B.Sc. in Electronics and Communications Engineering from the Benha Faculty of Engineering, Benha University (BU), Egypt, in 2008. He earned his MS in Communications and Computer Engineering from the Faculty of Engineering, BU, in 2013. Previously, he was an Assistant Lecturer at the Faculty of Engineering, BU. He is a Principal Embedded Engineer with the IoT and Networks Division at Orange Innovation Egypt, overseeing firmware architecture, development, and integration for advanced embedded systems and IoT solutions. His research and technical interests include software-defined radio, machine-to-machine communication, digital communication transceivers, low-power embedded systems, and IoT architectures.

**Khaled AlKady**, Khaled obtained his B.Sc. in Computer Science in 2005 from the Faculty of Computers and Information at Cairo University - Egypt. He later transitioned to the field of robotics and autonomous navigation systems, completing his M.Sc. in Autonomous Systems in 2019 at H-BRS University of Applied Sciences - Germany. Currently, he leads the Edge-AI and Embedded Systems unit at the Orange Innovation Egypt. His work in Edge-AI focuses on designing and optimizing deep learning solutions for edge devices with limited computational capabilities. The Embedded Systems side cover building ultra-low-power IoT solutions, mobile ad-hoc networks and algorithms for ambient backscattering communication in 4G networks.

**Amal Elkhodary**, Amal received her B.Sc. in engineering from the Faculty of Engineering, Suez Canal University, in 2021. After graduation, she completed the 9-month Wireless Communication Track program at the Information Technology Institute (ITI), Egypt. In 2023, she joined Orange Innovation, Egypt, where she has been engaged in research and development, focusing on backscattering tags and their application within 4G wireless communication systems. Her research interests include embedded systems, digital signal processing, and advancements in wireless communication technologies.

**Dinh-Thuy Phan-Huy**, [SM] received her degree in engineering from Supelec in 2001 and her Ph.D. in electronics and telecommunications from the National Institute of Applied Sciences (INSA) of Rennes, France, in 2015. In 2001, she joined France Telecom R&D (now Orange Innovation), France, as a research project manager. Her current research interests include wireless communications and 6G. She is an Orange Fellow.

**Ayman M. Hassan**, Obtained his B.Sc. in Telecommunications Engineering from Benha High Institute of Technology, Egypt, in 1993. He worked towards his M.Sc. and Ph.D. in Spread Spectrum Communications from Cairo University in 1998 and 2002, resp. He is now heading the IoT and Networks team at Orange Innovation Center in Cairo, where he is responsible for designing, deploying, and deploying Proof of Concepts for Telecom Solutions in AMEA countries. IoT solutions of interest include Utility Metering, Ultra-Low-Power nano-computers, and LORA-based IoT solutions. Research interests include Power Line Communication, Spread Spectrum, Ad-Hoc Networks, Backscattering and Military Communications. He is also working as an associate professor at Benha University and an adjunct faculty member at the American University in Cairo.